\documentclass[
    reprint,
    amsmath,
    amssymb,
    aps,
    prresearch,
    floatfix
]{revtex4-2}

\usepackage[T1]{fontenc}
\usepackage{graphicx}
\usepackage{xcolor}
\usepackage{upgreek}

\renewcommand{\textcolor}[2]{#2}
 
\begin{document}

\title{\textcolor{blue}{Terahertz Lattice Imprint of a Hidden Multipolar Ordering Transition in Gd$_2$Ru$_2$O$_7$}}

\author{Nicolas M. Kawahala}
\affiliation{Instituto de Física, Universidade de São Paulo, 05508-090, São Paulo, SP, Brazil}

\author{Rafael L. Sabainsk}
\altaffiliation{Present address: Institute of Physics, Goethe-University Frankfurt, Max-von-Laue-Str. 1, 60438 Frankfurt am Main, Germany.}
\affiliation{Instituto de Física, Universidade de São Paulo, 05508-090, São Paulo, SP, Brazil}

\author{Esteban Marulanda}
\affiliation{Instituto de Física, Universidade de São Paulo, 05508-090, São Paulo, SP, Brazil}

\author{Rafael S. Freitas}
\affiliation{Instituto de Física, Universidade de São Paulo, 05508-090, São Paulo, SP, Brazil}

\author{Felix G. G. Hernandez}
\email{Corresponding author: felixggh@if.usp.br}
\affiliation{Instituto de Física, Universidade de São Paulo, 05508-090, São Paulo, SP, Brazil}

\date{\today}

\begin{abstract}
\textcolor{blue}{Magnetic order carried by higher-rank multipoles produces no net moment and couples only weakly to conventional dipolar probes. Recently, the lattice has been proposed as an alternative detection channel for the associated ordering transitions. Here we report magneto-terahertz time-domain spectroscopy of the pyrochlore ruthenate Gd$_2$Ru$_2$O$_7$, in which a single spectrum resolves an
exchange-split Gd$^{3+}$ transition and an infrared-active optical phonon. The temperature and field dependence of the magnetic mode give two independent determinations of the internal exchange field at the rare-earth site, which yield a
magnitude of 5--6~T and a free-ion $g$-factor. The optical phonon, sensitive to spin-disorder scattering rather than to the static exchange field, exhibits a sharp anomaly in frequency and lifetime at 10~K that shifts to 30~K under 7~T, which we identify as an ordering transition of the Gd sublattice hidden from bulk thermodynamic probes. These results identify spin--phonon coupling as a route to rare-earth ordering transitions in multipolar magnets, one that is applicable to polycrystalline samples.}
\end{abstract}


\maketitle

\noindent\textit{Introduction}\,---\,Understanding how magnetic order couples to lattice degrees of freedom is a central problem in correlated electron systems~\cite{RevModPhys.83.471,10.21468/SciPostPhysCommRep.8}, where exchange interactions, local moments, and phonons can become strongly intertwined. Pyrochlore ruthenates containing magnetic rare-earth ions, $R_2$Ru$_2$O$_7$ ($R$ = rare earth), provide a particularly attractive platform to investigate this interplay owing to the coexistence of localized $4f$ moments and itinerant or partially localized Ru $4d$ magnetism~\cite{Greedan2001-hk,RevModPhys.82.53}. In these compounds, magnetic ordering of the Ru sublattice can generate an internal (molecular) exchange field that couples to the rare-earth ions, as demonstrated by the Zeeman-like splitting of the Ho$^{3+}$ crystal-electric-field (CEF) levels in Ho$_2$Ru$_2$O$_7$~\cite{PhysRevLett.93.076403,PhysRevB.108.054443,dzdf-hh4t}. Related CEF excitations have also been recently reported in Ho$_2$Zr$_2$O$_7$~\cite{rafael2025}. In contrast, Gd$_2$Ru$_2$O$_7$ exhibits comparatively weak CEF effects. Magnetic susceptibility and specific-heat measurements revealed a low-temperature Schottky-like anomaly, providing indirect evidence that Ru ordering splits the Gd$^{3+}$ ground-state multiplet through an internal exchange field and progressively polarizes the Gd moments~\cite{B110596P}. Although antiferromagnetic order of the Ru sublattice is established below $T_\textrm{N}\approx114$~K~\cite{PhysRevB.75.064426}, an independent phase transition associated with long-range ordering of the Gd moments has not been directly identified by bulk thermodynamic probes~\cite{PhysRevB.75.064426,B110596P,doi:10.7566/JPSJ.86.084708}. More recently, first-principles calculations have predicted a higher-order cluster-multipolar ground state involving the Gd sublattice in this material~\cite{Huebsch_2022}.

Across the family of rare-earth pyrochlores, multipolar magnetism has emerged as a central theme, with quadrupolar and octupolar degrees of freedom giving rise to hidden ordered states and quantum spin liquids~\cite{Sibille2020,Hayami2024,Poree2025}\textcolor{blue}{, whose fractionalized excitations have recently been predicted~\cite{Desrochers2024} and spectroscopically demarcated by field-dependent neutron scattering in Ce$_2$Zr$_2$O$_7$~\cite{Gao2026}. Because these higher-rank correlations couple only weakly to conventional dipolar probes, a parallel effort has developed around optical and lattice-based channels. For instance, multipolar symmetry breaking is predicted to generate a sizable dual-circular Raman optical activity in which a multipolar phonon plays an essential role~\cite{Watanabe2025}, an effect recently measured in the electric-toroidal-octupolar state of single-crystalline pyrite FeS$_2$~\cite{Suganuma2026}; octupolar order renders degenerate optical phonons pseudochiral, motivating helicity-resolved Raman as a probe of hidden order~\cite{Sutcliffe2026}; ferroic-octupolar order has been detected through symmetry-forbidden second-harmonic generation~\cite{Usachev2026}; and optically induced multipolar order is predicted to couple to the lattice and leave structural fingerprints~\cite{Banerjee2026}. These approaches establish the lattice as a viable channel for multipolar order, but they are largely nonlinear or polarization-resolved and therefore demand oriented single crystals. Complementary approaches sensitive to lattice responses and excitation dynamics in linear response are therefore increasingly important~\cite{Patri2020}.}

In this vein, 
anomalies in phonon frequencies and linewidths across magnetic phase transitions provide evidence for spin--lattice coupling in a broad range of correlated materials~\cite{6chp-2z42,PhysRevLett.94.137202,PhysRevB.96.054432,doi:10.1126/science.abk1121}, yet without direct access to the internal exchange field underlying these effects. In pyrochlore ruthenates, similar behavior has been reported in the nonmagnetic rare-earth site analogue Y$_2$Ru$_2$O$_7$, where anomalous spin dynamics and strong spin--phonon coupling have been observed~\cite{PhysRevB.77.020405,PhysRevB.69.214428}. 
\textcolor{blue}{What has not been established, however, is a route to the internal exchange field itself, its magnitude and its orientation, together with its imprint on a distinct lattice excitation across a magnetic ordering transition.} 
In this context, terahertz time-domain spectroscopy (THz-TDS) offers a unique opportunity to probe magnetic excitations and infrared-active optical phonons within the same spectral window~\cite{RevModPhys.83.471,phononsPbTe,phononsPbSnTe,Marulanda2026}. \textcolor{blue}{In rare-earth pyrochlores specifically, magneto-optical terahertz spectroscopy has resolved the field splitting of Tb$^{3+}$ crystal-field doublets in Tb$_2$Ti$_2$O$_7$ and shown that quadrupolar vibronic coupling is required to describe them~\cite{Amelin2020}, establishing the sensitivity of this spectral window to both rare-earth level structure and its coupling to the lattice.}
  
Here we use magneto-THz-TDS to directly access the low-energy electrodynamics of Gd$_2$Ru$_2$O$_7$. We identify a low-frequency excitation that emerges at low temperature and exhibits a strong field dependence, consistent with a transition between exchange-split Gd$^{3+}$ magnetic levels. From its temperature and field evolution, we obtain two independent spectroscopic determinations of the internal exchange field acting on the Gd sites, \textcolor{blue}{together with its orientation relative to the applied field}. In parallel, we resolve a strongly temperature- and field-dependent optical phonon whose frequency, linewidth, and spectral weight display comparably sharp anomalies at the established Ru-ordering temperature and at the temperature scale associated with Gd ordering. Tracking both excitations within a single spectroscopic framework, we show that the internal exchange field is simultaneously encoded in the lattice response, establishing a spectroscopy route to unresolved multipolar ordering transitions and spin--lattice coupling in correlated pyrochlores.

\

\noindent\textit{Results and Discussion}\,---\,Figure~1(a) shows the pyrochlore crystal structure of Gd$_2$Ru$_2$O$_7$, consisting of two interpenetrating magnetic sublattices of corner-sharing Gd$^{3+}$ and Ru$^{4+}$ tetrahedra. Figure~1(b) illustrates the Faraday-geometry magneto-THz-TDS configuration~\cite{koch_terahertz_2023,Baydin2021-je,THz_LA} employed to measure the pellet sample under external magnetic fields $B_\textrm{ext}$ up to 7~T and at temperatures between 1.6 and 200~K. The real optical conductivity spectra $\sigma_1(\nu)$ were extracted from the terahertz transmission data~\cite{Lloyd-Hughes2012-es,windowing,apertures}. Further experimental and analysis details are provided in the Supplemental Material~\cite{SM}.

The temperature dependence of $\sigma_1(\nu)$ at $B_\textrm{ext}=$ 0~T is shown in Fig.~1(c). Near 2.6~THz, a broad resonance-like feature persists across the entire temperature range, while a second contribution around 2~THz rapidly weakens upon cooling. Features in this range have been attributed to O--Gd--O optical phonons~\cite{ma17112571} or a magnetic spin gap of the Ru sublattice~\cite{PhysRevB.77.020405}. At 1.4~THz, a sharp resonance dominates the spectral weight and exhibits anomalies in its lineshape. The inset of Fig.~1(c) further reveals a much weaker mode around 0.15~THz emerging below 30~K. In the following, we focus on the two lower-frequency modes.

Figure~2(a) highlights the frequency range around the weaker mode ($\nu<$ 0.5~THz) and compares the corresponding response at $B_\textrm{ext}=$ 0~T and 7~T for representative temperatures. At 1.6~K, the resonance frequency for $B_\textrm{ext}=$ 7~T is nearly twice its zero-field value, establishing the magnetic origin of the mode, which remains clearly visible over the full temperature range. Moreover, the finite frequency of the mode at zero external field and its suppression above 30~K indicate the presence of an internal exchange field $B_\textrm{int}$ that becomes significant at low temperatures. This scale coincides with the broad Schottky-like anomaly reported in the specific heat of Gd$_2$Ru$_2$O$_7$~\cite{B110596P}, attributed to the splitting of the Gd$^{3+}$ multiplet by the molecular field of the Ru sublattice, suggesting that the observed excitation may originate from the same exchange-splitting mechanism. 

\begin{figure}
    \centering
    \includegraphics{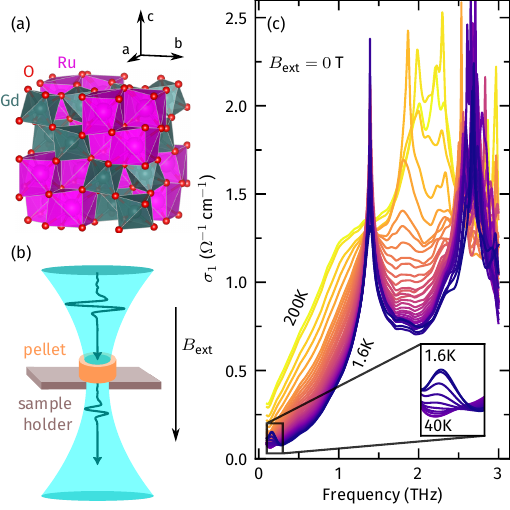}
    \caption{(a) Gd$_2$Ru$_2$O$_7$ crystal structure, rendered from data in the Materials Project~\cite{osti_1262381}. (b) Schematic of the Faraday-geometry magneto-THz-TDS setup near the sample position. (c) Real part of the optical conductivity at zero external magnetic field for temperatures between 1.6 (blue) and 200~K (yellow). \textit{Inset:} Magnified view of the low-frequency region, highlighting the emergence of a resonance near 0.15~THz at low temperatures.}
    \label{fig1}
\end{figure}

Lorentz-model fits~\cite{SM} of this mode yield the oscillator strength $\Omega_\textrm{m}$, resonance frequency $\nu_\textrm{m}$, and scattering time $\tau_\textrm{m}$ shown in Figs.~2(b--d). At $B_\textrm{ext}=$ 0~T, these parameters evolve monotonically with temperature, whereas under 7~T the mode exhibits two distinct regimes. Upon cooling from 200~K, the mode remains nearly unchanged down to 110~K before undergoing a marked hardening accompanied by substantial increases in both spectral weight and damping. The onset of this evolution coincides with $T_\textrm{N}$~\cite{PhysRevB.75.064426}, indicating that the excitation is directly sensitive to the development of Ru magnetic order.

\begin{figure*}
    \centering
    \includegraphics{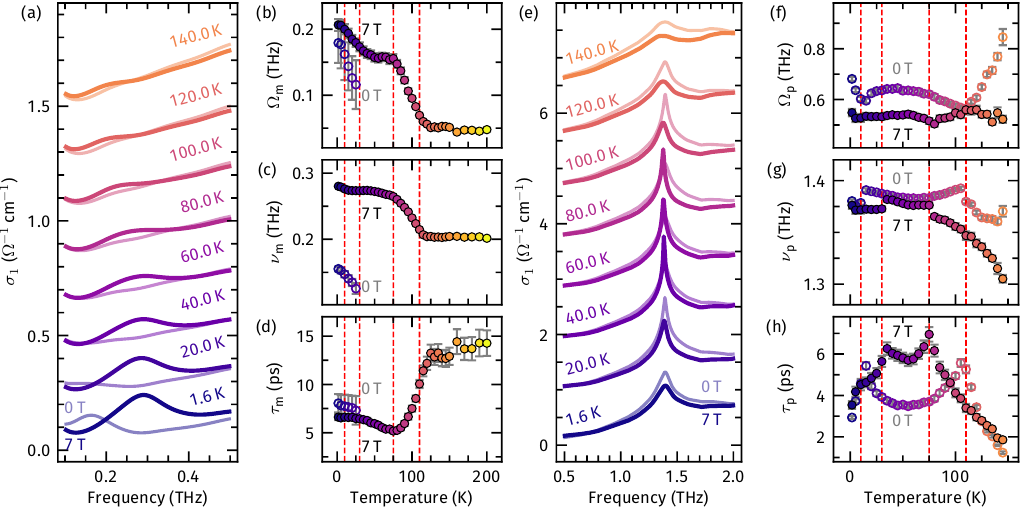}
    \caption{(a) Low-frequency optical conductivity showing a resonance associated with exchange-split Gd levels, for representative temperatures, measured at $B_\textrm{ext}=$ 0~T (shaded lines) and 7~T (solid lines). Temperature dependence of the fitted (b) oscillator strength, (c) resonance frequency, and (d) scattering time for $B_\textrm{ext}=$ 0~T (hollow markers) and 7~T (filled markers). Vertical dashed lines mark characteristic temperatures associated with Ru- and Gd-related anomalies discussed in the text, with values 10, 30, 75 and 110~K. (e) Optical conductivity in the phonon frequency range at the same external magnetic fields and temperatures shown in (a). (f)--(h) Temperature dependence of the corresponding phonon parameters. For clarity, spectra corresponding to different temperatures in (a) and (e) are vertically offset.}
    \label{fig2}
\end{figure*}

Below 75~K, the mode at $B_\textrm{ext}=$ 7~T enters the second temperature regime. While $\nu_\textrm{m}$ continues to evolve through a weak hardening and the damping is gradually reduced upon cooling, $\Omega_\textrm{m}$ exhibits a plateau between 75 and 40~K followed by a renewed increase. Although the monotonic evolution of $\Omega_\textrm{m}$ and $\nu_\textrm{m}$ observed at $B_\textrm{ext}=$ 0~T is consistent with the progressive polarization of the Gd moments induced by the molecular field established by Ru ordering, the results under $B_\textrm{ext}=$ 7~T cannot be explained by this crossover alone. The renewed increase of $\Omega_\textrm{m}$ and the slightly stronger hardening of $\nu_\textrm{m}$ below 40~K suggest a further enhancement of $B_\textrm{int}$ associated with the onset of a distinct low-temperature ordered state. \textcolor{blue}{Such an ordering is expected to involve the Gd sublattice in this temperature range \cite{RevModPhys.82.53}. $^{155}$Gd M\"ossbauer spectroscopy shows the Gd moment growing steeply below about 40~K, although no phase transition was resolved between $T_\textrm{N}$ and 4.2~K~\cite{PhysRevB.75.064426}.} Recent first-principles calculations further predict this state to adopt a higher-order cluster-multipolar configuration in Gd$_2$Ru$_2$O$_7$~\cite{Huebsch_2022}, which is expected to remain largely elusive to conventional thermodynamic probes~\cite{B110596P}.

Collectively, the observations in Figs.~2(a--d) identify this mode as a transition within the exchange-split $^8S_{7/2}$ multiplet of Gd$^{3+}$, governed by an effective field comprising external and internal exchange contributions. Furthermore, they provide a spectroscopic measure of $B_\textrm{int}$ acting on the rare-earth site. At $B_\textrm{ext}=$ 7~T and 200~K, well above $T_\textrm{N}$ where $B_\textrm{int}$ is absent, the measured $\nu_\textrm{m}$ corresponds to a splitting $\Delta_\textrm{ext}=$ 0.202(1)~THz, establishing the calibration $\Delta/B=$ 0.029(1)~THz/T. \textcolor{blue}{Since $\Delta=g\mu_\textrm{B}B/h$, this calibration constitutes a direct determination of the ground-multiplet $g$-factor, $g=h(\Delta/B)/\mu_\textrm{B}=$ 2.07(7), in agreement with the value $g\approx2$ expected for the $L=0$, $S=7/2$ ground state of the half-filled $4f^7$ shell of Gd$^{3+}$, and confirming the magnetic-dipole character of the transition.} Hence, at 1.6~K and zero applied field, the measured splitting $\Delta_\textrm{int}=$ 0.155(3)~THz yields $B_\textrm{int}=$ 5.4(2)~T. \textcolor{blue}{We compare this value with the molecular field expected at the Gd site. Analysis of $^{99}$Ru and $^{155}$Gd M\"ossbauer spectra gives the molecular-field coefficients $\lambda_\textrm{Ru-Gd}=\pm14(1)$~kOe/$\mu_\textrm{B}$ and $\lambda_\textrm{Gd-Gd}=-4.9(6)$~kOe/$\mu_\textrm{B}$~\cite{PhysRevB.75.064426}. With the saturated Gd moment of 7~$\mu_\textrm{B}$ and Ru moments in the reported range 1.2--2.6~$\mu_\textrm{B}$, the Ru and Gd sublattices contribute 1.7--3.6~T and 3.4~T, respectively, so that a field of 5--7~T is expected at the Gd site. Our spectroscopic determination agrees with this hyperfine estimate, and the comparison of magnitudes is independent of the orientation of the crystallites.}

At $B_\textrm{ext}=$ 7~T and below $T_\textrm{N}$, where the effective field comprises external and internal contributions, the corresponding splitting satisfies $\Delta^2 = \Delta_{\rm ext}^2 + \Delta_{\rm int}^2 + 2\Delta_{\rm ext}\Delta_{\rm int}\cos\theta$. The measured splitting $\Delta=$ 0.280(1)~THz at 1.6~K, together with the extracted values of $\Delta_\textrm{ext}$ and $\Delta_\textrm{int}$, yields $\theta=$ 78(1)$^\circ$, corresponding to an internal field nearly transverse to the applied field. \textcolor{blue}{Because the sample is polycrystalline, $\theta$ is an effective angle
between $B_\textrm{int}$ and the laboratory field, and not a crystallographic one. An internal field oriented at random with respect to $B_\textrm{ext}$ would give $\langle\cos\theta\rangle=0$ and a splitting of 0.255~THz at 7~T, well below the measured value. The internal field is therefore not isotropically distributed relative
to the applied field. For context, we note that previous measurements of the angle enclosed by the Ru moments and the local [111] direction reported a value of $72^\circ$~\cite{PhysRevB.75.064426}, while more recent calculations predicted $82^\circ$ for the Ru and $80^\circ$ for the Gd sublattices~\cite{Huebsch_2022}.}

We next examine the dominant mode near 1.4~THz. Figure~2(e) shows the optical conductivity spectra in the range 0.5--2.0~THz for the same temperatures and fields as in Fig.~2(a). Lorentz-model fits~\cite{SM} yield $\Omega_\textrm{p}$, $\nu_\textrm{p}$, and $\tau_\textrm{p}$ shown in Figs.~2(f--h). The large spectral weight, persistence well above $T_\textrm{N}$ even at $B_\textrm{ext}=$ 0~T, and weak field dependence of $\nu_\textrm{p}$ identify this resonance as an infrared-active optical phonon. All three parameters exhibit pronounced anomalies at the characteristic temperatures identified above, indicating that the phonon dynamics are intimately coupled to the underlying magnetic energy scales.

At $B_\textrm{ext}=$ 0~T, $\nu_\textrm{p}$ hardens at 110~K and softens at 10~K, while $\tau_\textrm{p}$ exhibits pronounced maxima at both temperatures. Since the phonon linewidth is sensitive to spin-disorder scattering, which is suppressed as magnetic correlations develop, each maximum in $\tau_\textrm{p}$ is consistent with an abrupt reduction of spin fluctuations accompanying the onset of magnetic order, rather than a gradual crossover. Accordingly, the high-temperature anomaly coincides with the well-established Néel transition of the Ru sublattice, while its displacement to 75~K under $B_\textrm{ext}=$ 7~T is consistent with a field-induced suppression of Ru antiferromagnetic order.

Likewise, the comparably sharp anomaly near 10~K occurs at the temperature scale expected for Gd ordering~\cite{PhysRevB.75.064426,Huebsch_2022}, while its shift to 30~K under $B_\textrm{ext}=$ 7~T coincides with the renewed evolution of the exchange-split Gd$^{3+}$ excitation discussed above. The absence of a distinct zero-field counterpart in the magnetic excitation is consistent with the broad Schottky-like evolution of the exchange-split Gd multiplet~\cite{B110596P}, which spans this range and may obscure a narrower transition in both the magnetic and bulk thermodynamic responses. \textcolor{blue}{This 10~K anomaly is distinct from the crossover near 40~K below which the Gd hyperfine field rises steeply~\cite{PhysRevB.75.064426}. That crossover reflects the growth of the induced Gd moment, which is already close to saturation at 10~K.}

\textcolor{blue}{Three effects account for the absence of a signature in earlier work. The ordered Gd configuration, reported as a 32-pole~\cite{PhysRevB.75.064426} and predicted to be cluster-multipolar~\cite{Huebsch_2022}, carries no net moment for magnetization or dipolar magnetic scattering to detect. The asymmetry parameter of the electric-field gradient at the Gd site vanishes, so that $^{155}$Gd M\"ossbauer spectroscopy determines the magnitude of the hyperfine field and its polar angle but is insensitive to the moment directions within the plane perpendicular to the local threefold axis~\cite{PhysRevB.75.064426}, where the multipolar degree of freedom resides. In the specific heat, the magnetic entropy of the Schottky anomaly, 31.7~J~mol$^{-1}$~K$^{-1}$, is nearly eight times that released at the Ru ordering transition, 4.1~J~mol$^{-1}$~K$^{-1}$~\cite{B110596P}, so that a comparable anomaly of the Gd sublattice would be a small perturbation on a much larger smooth background.} The phonon therefore provides independent spectroscopic evidence for an otherwise elusive Gd-ordering transition.

\begin{figure*}
    \centering
    \includegraphics{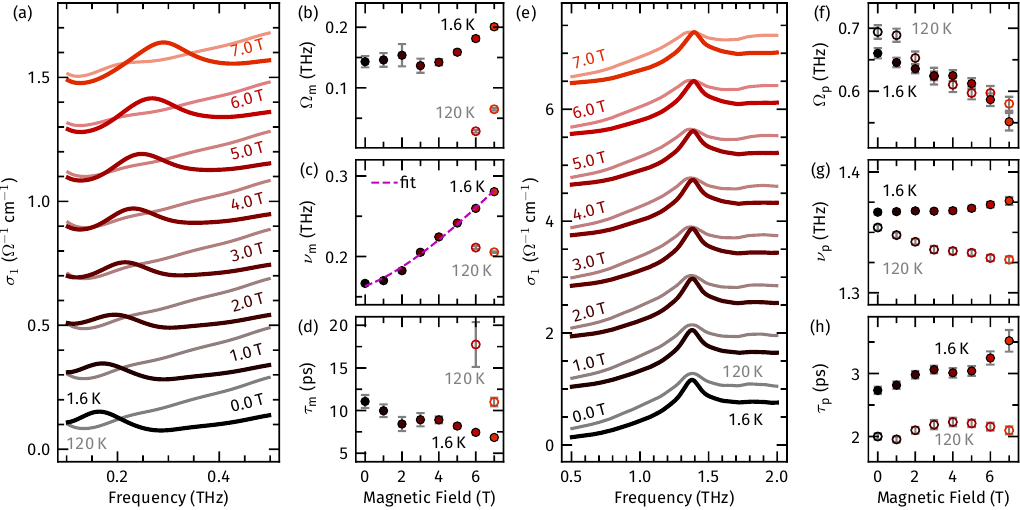}
    \caption{(a) Low-frequency optical conductivity at 1.6~K (solid lines) and 120~K (shaded lines) for external magnetic fields from 0 to 7~T, showing a field-induced shift and enhancement of the Gd-related resonance. External magnetic-field dependence of the extracted (b) spectral weight, (c) resonance frequency, and (d) scattering time at 1.6~K (filled markers) and 120~K (hollow markers). In (c), the dashed line is a fit including internal and external field contributions. (e) Optical conductivity in the phonon frequency range at the same external magnetic fields and temperatures shown in (a). (f)--(h) Magnetic-field dependence of the phonon parameters. For clarity, spectra corresponding to different magnetic fields in (a) and (e) are vertically offset.}
    \label{fig3}
\end{figure*}

To further examine the field dependence of both excitations, we measured the optical conductivity as a function of $B_\textrm{ext}$ at 1.6 and 120~K, corresponding to temperatures below and above $T_\textrm{N}$, respectively. Figure~3 compares these spectra in the same frequency ranges analyzed in Fig.~2, together with the field evolution of the fitted Lorentz-model parameters~\cite{SM}. At 1.6~K, the magnetic excitation [Figs.~3(a--d)] exhibits a pronounced field-induced hardening, whereas at 120~K the resonance becomes observable only above $B_\textrm{ext}=$ 6~T, consistent with the external field reaching the same energy scale provided by $B_\textrm{int}$ at low temperatures. 

The field dependence of $\nu_\textrm{m}$ is well captured by the Zeeman-like splitting $\Delta=g\mu_\textrm{B}B_\textrm{eff}/h$, with $B_\textrm{eff}^2=B_\textrm{ext}^2+B_\textrm{int}^2+2B_\textrm{ext}B_\textrm{int}\cos\theta$, as indicated by the dashed curve in Fig.~3(c). This fit yields $g=$ 1.86(17), $B_\textrm{int}=$ 6.3(5)~T, and $\theta=$ 70(8)$^\circ$. The extracted $g$-factor agrees within uncertainty \textcolor{blue}{both} with the value of 2 expected for isolated Gd$^{3+}$ moments \textcolor{blue}{and with the independent determination $g=$ 2.07(7) obtained above from the high-temperature calibration}, while $B_\textrm{int}$ and $\theta$ remain consistent with the values obtained from the temperature-dependent analysis. \textcolor{blue}{We emphasize that these determinations are extracted from disjoint data sets, a temperature sweep at fixed $B_\textrm{ext}=$ 7~T and a field sweep at fixed $T=$ 1.6~K, and share no fitted parameter, so that their mutual agreement provides internal confirmation of both the assignment of the mode and the extraction procedure.} The field evolution of $\Omega_\textrm{m}$ and $\tau_\textrm{m}$ is likewise compatible with the behavior discussed above, showing a moderate increase in spectral weight and damping.

The optical phonon exhibits a markedly different field response depending on temperature, as shown in Figs.~3(e--h). At 120~K, $\nu_\textrm{p}$ softens slightly with increasing $B_\textrm{ext}$, whereas at 1.6~K it remains nearly unchanged, showing only a weak hardening. This contrasting behavior suggests that, above $T_\textrm{N}$, the applied field modifies the magnetic correlations sufficiently to reduce the magnetic contribution to the lattice restoring force through spin--phonon coupling. Below $T_\textrm{N}$, by contrast, the magnetic correlations are already well established, so $B_\textrm{ext}$ produces only minor changes to the ordered spin configuration, resulting in a nearly field-independent phonon frequency.

The field evolution of $\tau_\textrm{p}$ further supports this interpretation. While the phonon lifetime increases only moderately with $B_\textrm{ext}$ [Fig.~3(h)], the much stronger linewidth narrowing observed upon cooling through the magnetic transitions [Fig.~2(h)] demonstrates that the long-range magnetic ordering is considerably more effective than the applied field in suppressing spin-disorder scattering. Together, the temperature- and field-dependent measurements show that the phonon dynamics are governed by the magnetic energy scales of both sublattices, providing further evidence for strong spin--phonon coupling in Gd$_2$Ru$_2$O$_7$. \textcolor{blue}{That a lattice mode should encode multipolar magnetic order is theoretically expected, as magnetic octupoles couple linearly to degenerate optical phonons and render them pseudochiral~\cite{Sutcliffe2026}, and multipolar order induced in correlated insulators generically produces measurable lattice distortions~\cite{Banerjee2026}.}

\

\noindent\textit{Conclusions}\,---\,In this work, terahertz time-domain spectroscopy was used to investigate the low-energy electrodynamics of Gd$_2$Ru$_2$O$_7$ under applied magnetic fields. A low-frequency magnetic excitation associated with exchange-split Gd$^{3+}$ levels provides direct spectroscopic access to the internal exchange field acting on the rare-earth site. Independent temperature- and field-dependent analyses yield $B_\textrm{int}\approx$ 5--6~T at low temperature, \textcolor{blue}{together with two mutually consistent determinations of the $g$-factor, $g=$ 2.07(7) and $g=$ 1.86(17), both in agreement with the value expected for the $L=0$ ground state of isolated Gd$^{3+}$ moments.} \textcolor{blue}{The magnitude agrees with the molecular field expected from the published $^{155}$Gd M\"ossbauer coefficients~\cite{PhysRevB.75.064426}, which does not depend on crystallite orientation.} The temperature evolution of this excitation further separates the onset of Ru order near $T_\textrm{N}$ from a renewed low-temperature enhancement associated with Gd ordering.

Simultaneously, we showed that an infrared-active optical phonon exhibits pronounced anomalies in its frequency, lifetime, and oscillator strength at the characteristic magnetic energy scales of both sublattices. While the high-temperature anomaly follows the established Ru ordering transition, the comparably sharp low-temperature anomaly identifies a genuine ordering transition of the Gd sublattice rather than a purely gradual polarization crossover. \textcolor{blue}{Three effects keep this transition out of reach of conventional probes.} \textcolor{blue}{The cluster-multipolar configuration predicted for the Gd sublattice~\cite{Huebsch_2022} has a vanishing cluster dipole on each tetrahedron, so it carries no net moment for magnetization or dipolar magnetic scattering to detect.} \textcolor{blue}{The vanishing asymmetry parameter of the electric-field gradient at the Gd site renders $^{155}$Gd M\"ossbauer spectroscopy insensitive to the in-plane moment directions~\cite{PhysRevB.75.064426}.} \textcolor{blue}{In the specific heat, a narrow transition anomaly falls on top of the broad Schottky background of the exchange-split Gd multiplet, which covers the same temperature range~\cite{B110596P}. With $\Delta_\textrm{int}=$ 0.155(3)~THz, the seven intervals of the $^8S_{7/2}$ multiplet extend over $\approx52$~K,} \textcolor{blue}{and that background carries nearly eight times the entropy released at the Ru ordering transition~\cite{B110596P}.} The phonon, by contrast, is directly sensitive to changes in local spin correlations through spin--phonon coupling, offering a complementary route to detect this transition independently of the probes that have so far not resolved it.

\textcolor{blue}{The present approach is complementary to the routes currently being developed to detect multipolar order. Nonlinear and polarization-resolved optical probes, such as second-harmonic generation~\cite{Usachev2026}, dual-circular Raman optical activity~\cite{Watanabe2025,Suganuma2026}, and helicity-resolved Raman of pseudochiral phonons~\cite{Sutcliffe2026}, derive their selectivity from symmetry and therefore require oriented single crystals, while neutron spectroscopy of dipolar-octupolar pyrochlores~\cite{Desrochers2024,Gao2026} requires both large crystals and weakly absorbing rare-earth ions. The measurement reported here instead operates in linear response on a polycrystalline sample, and derives its selectivity from the magnetic-field dependence of two excitations of different character recorded within a single spectrum.}

By resolving magnetic and lattice excitations within the same spectral window, our results establish a direct spectroscopic connection between internal exchange fields, a proposed multipolar order on the Gd sublattice, and spin--phonon coupling in a rare-earth pyrochlore ruthenate. More broadly, this work demonstrates that terahertz spectroscopy, through its sensitivity to lattice dynamics, offers a route to detect magnetic ordering transitions that remain hidden from conventional thermodynamic probes, opening new opportunities to investigate hidden and multipolar order in frustrated magnetic materials.

\

\noindent\textit{Acknowledgments}\,---\,We acknowledge financial support from the São Paulo Research Foundation (FAPESP) under Grant Nos. 2021/12470-8, 2023/04245-0, and 2023/16742-8. F.G.G.H. and R.S.F. acknowledge financial support from the National Council for Scientific and Technological Development (CNPq) under Grant Nos. 306550/2023-7 and 302689/2025-7, respectively. E.M. acknowledges financial support from Grant No. 88887.007580/2024-00 of the Coordenação de Aperfeiçoamento de Pessoal de Nível Superior (CAPES). Additional support from the INCT Advanced Quantum Materials project funded by CNPq (Grant No. 408766/2024-7), FAPESP (Grant No. 2025/27091-3), and CAPES is also gratefully acknowledged.

\

\noindent\textit{Data availability}\,---\,The data that support the findings
of this article are not publicly available. The data are
available from the authors upon reasonable request.

\bibliography{bibl.bib}

@Article{ma17112571,
AUTHOR = {Kraidy, Assohoun Fulgence and Yapi, Abé Simon and El Marssi, Mimoun and Penton Madrigal, Arbelio and Gagou, Yaovi},
TITLE = {{Structural Refinement and Optoelectrical Properties of Nd2Ru2O7 and Gd2Ru2O7 Pyrochlore Oxides for Photovoltaic Applications}},
JOURNAL = {Materials},
VOLUME = {17},
YEAR = {2024},
NUMBER = {11},
ARTICLE-NUMBER = {2571},
URL = {https://www.mdpi.com/1996-1944/17/11/2571},
PubMedID = {38893833},
ISSN = {1996-1944},
DOI = {10.3390/ma17112571}
}

@article{PhysRevB.75.064426,
  title = {{Bulk magnetic measurements and $^{99}\mathrm{Ru}$ and $^{155}\mathrm{Gd}$ M\"ossbauer spectroscopies of ${\mathrm{Gd}}_{2}{\mathrm{Ru}}_{2}{\mathrm{O}}_{7}$}},
  author = {Gurgul, J. and Rams, M. and \ifmmode \acute{S}\else \'{S}\fi{}wi\k{a}tkowska, \ifmmode \dot{Z}\else \.{Z}\fi{}. and Kmie\ifmmode \acute{c}\else \'{c}\fi{}, R. and Tomala, K.},
  journal = {Phys. Rev. B},
  volume = {75},
  issue = {6},
  pages = {064426},
  numpages = {8},
  year = {2007},
  month = {Feb},
  publisher = {American Physical Society},
  doi = {10.1103/PhysRevB.75.064426},
  url = {https://link.aps.org/doi/10.1103/PhysRevB.75.064426}
}

@ARTICLE{Greedan2001-hk,
  title     = "Geometrically frustrated magnetic materials",
  author    = "Greedan, John E",
  journal   = "J. Mater. Chem.",
  publisher = "Royal Society of Chemistry (RSC)",
  volume    =  11,
  number    =  1,
  pages     = "37--53",
  year      =  2001
}

@article{PhysRevLett.93.076403,
  title = {{Magnetic Ordering in the Spin-Ice Candidate ${\mathrm{H}\mathrm{o}}_{2}{\mathrm{R}\mathrm{u}}_{2}{\mathrm{O}}_{7}$}},
  author = {Wiebe, C. R. and Gardner, J. S. and Kim, S.-J. and Luke, G. M. and Wills, A. S. and Gaulin, B. D. and Greedan, J. E. and Swainson, I. and Qiu, Y. and Jones, C. Y.},
  journal = {Phys. Rev. Lett.},
  volume = {93},
  issue = {7},
  pages = {076403},
  numpages = {4},
  year = {2004},
  month = {Aug},
  publisher = {American Physical Society},
  doi = {10.1103/PhysRevLett.93.076403},
  url = {https://link.aps.org/doi/10.1103/PhysRevLett.93.076403}
}

@article{dzdf-hh4t,
  title = {{Ferromagnetic fragmented state in the pyrochlore ${\mathrm{Ho}}_{2}{\mathrm{Ru}}_{2}{\mathrm{O}}_{7}$}},
  author = {Museur, F. and Robert, J. and Morineau, F. and Bujault, N. and Simonet, V. and Pachoud, E. and Hadj-Azzem, A. and Colin, C. V. and Mangin-Thro, L. and Manuel, P. and Stewart, J. R. and Holdsworth, P. C. W. and Lhotel, E.},
  journal = {Phys. Rev. B},
  volume = {113},
  issue = {6},
  pages = {L060406},
  numpages = {7},
  year = {2026},
  month = {Feb},
  publisher = {American Physical Society},
  doi = {10.1103/dzdf-hh4t},
  url = {https://link.aps.org/doi/10.1103/dzdf-hh4t}
}

@article{PhysRevB.77.020405,
  title = {{From cooperative paramagnetism to N\'eel order in ${\mathrm{Y}}_{2}{\mathrm{Ru}}_{2}{\mathrm{O}}_{7}$: Neutron scattering measurements}},
  author = {van Duijn, J. and Hur, N. and Taylor, J. W. and Qiu, Y. and Huang, Q. Z. and Cheong, S.-W. and Broholm, C. and Perring, T. G.},
  journal = {Phys. Rev. B},
  volume = {77},
  issue = {2},
  pages = {020405},
  numpages = {4},
  year = {2008},
  month = {Jan},
  publisher = {American Physical Society},
  doi = {10.1103/PhysRevB.77.020405},
  url = {https://link.aps.org/doi/10.1103/PhysRevB.77.020405}
}

@article{Huebsch_2022,
doi = {10.1088/1361-648X/ac513c},
url = {https://doi.org/10.1088/1361-648X/ac513c},
year = {2022},
month = {mar},
publisher = {IOP Publishing},
volume = {34},
number = {19},
pages = {194003},
author = {Huebsch, M-T and Nomura, Y and Sakai, S and Arita, R},
title = {Magnetic structures and electronic properties of cubic-pyrochlore ruthenates from first principles},
journal = {Journal of Physics: Condensed Matter}
}

@Article{B110596P,
author ="Taira, Nobuyuki and Wakeshima, Makoto and Hinatsu, Yukio",
title  ="{Magnetic susceptibility and specific heat studies on heavy rare earth ruthenate pyrochlores R2Ru2O7 (R = Gd–Yb)}",
journal  ="J. Mater. Chem.",
year  ="2002",
volume  ="12",
issue  ="5",
pages  ="1475-1479",
publisher  ="The Royal Society of Chemistry",
doi  ="10.1039/B110596P",
url  ="http://dx.doi.org/10.1039/B110596P"}

@article{RevModPhys.82.53,
  title = {Magnetic pyrochlore oxides},
  author = {Gardner, Jason S. and Gingras, Michel J. P. and Greedan, John E.},
  journal = {Rev. Mod. Phys.},
  volume = {82},
  issue = {1},
  pages = {53--107},
  numpages = {0},
  year = {2010},
  month = {Jan},
  publisher = {American Physical Society},
  doi = {10.1103/RevModPhys.82.53},
  url = {https://link.aps.org/doi/10.1103/RevModPhys.82.53}
}

@article{PhysRevB.69.214428,
  title = {{Strong spin-phonon coupling in the geometrically frustrated pyrochlore ${\mathrm{Y}}_{2}{\mathrm{Ru}}_{2}{\mathrm{O}}_{7}$}},
  author = {Lee, J. S. and Noh, T. W. and Bae, J. S. and Yang, In-Sang and Takeda, T. and Kanno, R.},
  journal = {Phys. Rev. B},
  volume = {69},
  issue = {21},
  pages = {214428},
  numpages = {5},
  year = {2004},
  month = {Jun},
  publisher = {American Physical Society},
  doi = {10.1103/PhysRevB.69.214428},
  url = {https://link.aps.org/doi/10.1103/PhysRevB.69.214428}
}

@article{doi:10.7566/JPSJ.86.084708,
author = {Okamura ,Takuma and Okazaki ,Ryuji and Taniguchi ,Hiroki and Yasui ,Yukio and Terasaki ,Ichiro},
title = {{Magnetoelectric Coupling in the Pyrochlore Ruthenate Gd2Ru2O7}},
journal = {Journal of the Physical Society of Japan},
volume = {86},
number = {8},
pages = {084708},
year = {2017},
doi = {10.7566/JPSJ.86.084708}
}

@article{6chp-2z42,
  title = {{Short-range spin order and spin-phonon coupling in ${\mathrm{NaCrO}}_{2}$ revealed by terahertz spectra}},
  author = {Huang, Biwen and Hao, Pengli and Sheng, Zhigao},
  journal = {Phys. Rev. B},
  volume = {113},
  issue = {9},
  pages = {094435},
  numpages = {7},
  year = {2026},
  month = {Mar},
  publisher = {American Physical Society},
  doi = {10.1103/6chp-2z42},
  url = {https://link.aps.org/doi/10.1103/6chp-2z42}
}

@article{PhysRevLett.94.137202,
  title = {{Probing Spin Correlations with Phonons in the Strongly Frustrated Magnet ${\mathrm{ZnCr}}_{2}{\mathrm{O}}_{4}$}},
  author = {Sushkov, A. B. and Tchernyshyov, O. and II, W. Ratcliff and Cheong, S. W. and Drew, H. D.},
  journal = {Phys. Rev. Lett.},
  volume = {94},
  issue = {13},
  pages = {137202},
  numpages = {4},
  year = {2005},
  month = {Apr},
  publisher = {American Physical Society},
  doi = {10.1103/PhysRevLett.94.137202},
  url = {https://link.aps.org/doi/10.1103/PhysRevLett.94.137202}
}

@article{PhysRevB.96.054432,
  title = {{Terahertz dielectric analysis and spin-phonon coupling in multiferroic ${\mathrm{GeV}}_{4}{\mathrm{S}}_{8}$}},
  author = {Warren, Matthew T. and Pokharel, G. and Christianson, A. D. and Mandrus, D. and Vald\'es Aguilar, R.},
  journal = {Phys. Rev. B},
  volume = {96},
  issue = {5},
  pages = {054432},
  numpages = {7},
  year = {2017},
  month = {Aug},
  publisher = {American Physical Society},
  doi = {10.1103/PhysRevB.96.054432},
  url = {https://link.aps.org/doi/10.1103/PhysRevB.96.054432}
}

@article{
doi:10.1126/science.abk1121,
author = {Evgeny A. Mashkovich  and Kirill A. Grishunin  and Roman M. Dubrovin  and Anatoly K. Zvezdin  and Roman V. Pisarev  and Alexey V. Kimel },
title = {Terahertz light–driven coupling of antiferromagnetic spins to lattice},
journal = {Science},
volume = {374},
number = {6575},
pages = {1608-1611},
year = {2021},
doi = {10.1126/science.abk1121}}

@article{PhysRevB.108.054443,
  title = {{Linear scaling relationship of N\'eel temperature and dominant magnons in pyrochlore ruthenates}},
  author = {Lee, Jae Hyuck and Wulferding, Dirk and Kim, Junkyoung and Song, Dongjoon and Park, Seung Ryong and Kim, Changyoung},
  journal = {Phys. Rev. B},
  volume = {108},
  issue = {5},
  pages = {054443},
  numpages = {8},
  year = {2023},
  month = {Aug},
  publisher = {American Physical Society},
  doi = {10.1103/PhysRevB.108.054443},
  url = {https://link.aps.org/doi/10.1103/PhysRevB.108.054443}
}

@article{koch_terahertz_2023,
    title = {Terahertz time-domain spectroscopy},
    volume = {3},
    doi = {10.1038/s43586-023-00232-z},
    number = {1},
    journal = {Nature Reviews Methods Primers},
    author = {Koch, Martin and Mittleman, Daniel M. and Ornik, Jan and Castro-Camus, Enrique},
    year = {2023},
    pages = {1--14},
}

@ARTICLE{Baydin2021-je,
  title     = "Time-domain terahertz spectroscopy in high magnetic fields",
  author    = "Baydin, Andrey and Makihara, Takuma and Peraca, Nicolas Marquez and Kono, Junichiro",
  journal   = "Front. Optoelectron.",
  volume    =  14,
  number    =  1,
  pages     = "110--129",
  month     =  mar,
  year      =  2021,
}

@article{THz_LA,
    title = {{Terahertz and mm-Wave Research in Latin America}},
    doi = {10.1007/s10762-026-01131-6},
    author = {Castro-Camus, Enrique and Ferrusca, Daniel and Carpenter, John and Hughes, David and Qureshi, Naser and Freitas, Raul O. and Strupiechonski, Elodie and Oblitas, Jimy and Alfaro-Gomez, Mariana and Gomez-Sepulveda, Alma Montserrat and Hernandez, Felix G. G. and Sanjuan, Federico and Ortiz-Martinez, Monica},
    year = {2026},
    journal = {Journal of Infrared, Millimeter, and Terahertz Waves},
    volume = {47},
    pages = {27},
}

@article{Marulanda2026,
  author = {E. Marulanda and M. Dutra and N. M. Kawahala and E. D. Stefanato and G. G. Vasques and J. Munevar and M. A. Avila and F. G. G. Hernandez},
  title = {{Colossal Terahertz Magnetoresistance from Magnetic Polarons in EuZn$_2$P$_2$}},
  year = {2026},
  eprint = {2603.21423},
  archivePrefix = {arXiv},
  journal = {},
  primaryClass = {cond-mat.str-el}
}

@article{phononsPbTe,
    title = {Magnetic Control of Soft Chiral Phonons in {PbTe}},
    author = {Baydin, Andrey and Hernandez, Felix G. G. and Rodriguez-Vega, Martin and Okazaki, Anderson K. and Tay, Fuyang and Noe, G. Timothy and Katayama, Ikufumi and Takeda, Jun and Nojiri, Hiroyuki and Rappl, Paulo H. O. and Abramof, Eduardo and Fiete, Gregory A. and Kono, Junichiro},
    journal = {Phys. Rev. Lett.},
    volume = {128},
    issue = {7},
    pages = {075901},
    numpages = {6},
    year = {2022},
    month = {2},
    publisher = {American Physical Society},
    doi = {10.1103/PhysRevLett.128.075901}
}

@article{phononsPbSnTe,
    title = {Observation of interplay between phonon chirality and electronic band topology},
    volume = {9},
    issn = {2375-2548},
    doi = {10.1126/sciadv.adj4074},
    number = {50},
    journal = {Science Advances},
    author = {Hernandez, Felix G. G. and Baydin, Andrey and Chaudhary, Swati and Tay, Fuyang and Katayama, Ikufumi and Takeda, Jun and Nojiri, Hiroyuki and Okazaki, Anderson K. and Rappl, Paulo H. O. and Abramof, Eduardo and Rodriguez-Vega, Martin and Fiete, Gregory A. and Kono, Junichiro},
    month = {12},
    year = {2023},
    pages = {eadj4074},
}

@article{windowing,
    title = {Windowing in terahertz time-domain spectroscopy: resolving resonances in thin-film samples},
    doi = {10.1007/s10762-025-01092-2},
    author = {Marulanda, Esteban and Costa, Fernanda L. and Kawahala, Nicolas M. and Hernandez, Felix G. G.},
    year = {2025},
    journal = {Journal of Infrared, Millimeter, and Terahertz Waves},
    volume = {46},
    number = {10},
    pages = {75},
}

@article{apertures,
      title={{Optimizing Aperture Geometry in THz-TDS for Accurate Spectroscopy of Quantum Materials}}, 
      author={Laura O. Dias and Eduardo D. Stefanato and Nicolas M. Kawahala and Felix G. G. Hernandez},
      year={2026},
      volume={56},
      number={1},
      pages={35},
      journal={Brazilian Journal of Physics},
      doi={10.1007/s13538-025-01973-w}
}

@article{Lloyd-Hughes2012-es,
  title     = {A review of the terahertz conductivity of bulk and
               nano-materials},
  author    = {Lloyd-Hughes, James and Jeon, Tae-In},
  journal   = {Journal of Infrared, Millimeter, and Terahertz Waves},
  volume    =  {33},
  number    =  {9},
  pages     = {871},
  year      =  {2012},
  doi={10.1007/s10762-012-9905-y}
}

@article{RevModPhys.83.471,
  title = {Electrodynamics of correlated electron materials},
  author = {Basov, D. N. and Averitt, Richard D. and van der Marel, Dirk and Dressel, Martin and Haule, Kristjan},
  journal = {Rev. Mod. Phys.},
  volume = {83},
  issue = {2},
  pages = {471--541},
  numpages = {0},
  year = {2011},
  month = {Jun},
  publisher = {American Physical Society},
  doi = {10.1103/RevModPhys.83.471}
  }

@misc{SM,
  note = {See Supplemental Material at [URL will be inserted by publisher] for additional experimental and analysis details.}
}

@article{rafael2025,
  title = {{Excitation spectrum and low-temperature magnetism in the disordered defect-fluorite ${\mathrm{Ho}}_{2}{\mathrm{Zr}}_{2}{\mathrm{O}}_{7}$}},
  author = {Oliveira Silva, P. L. and Ramon, J. G. A. and Peçanha-Antonio, Viviane and Guidi, Tatiana and Gardner, J. S. and Fang, Chun Sheng and Freitas, R. S.},
  journal = {Phys. Rev. B},
  volume = {113},
  issue = {21},
  pages = {214423},
  numpages = {10},
  year = {2026},
  month = {Jun},
  publisher = {American Physical Society},
  doi = {10.1103/rf1x-8swz},
  url = {https://link.aps.org/doi/10.1103/rf1x-8swz}
}

@Article{10.21468/SciPostPhysCommRep.8,
	title={{The future of the correlated electron problem}},
	author={A. Alexandradinata and N. P. Armitage and Andrey Baydin and Wenli Bi and Yue Cao and Hitesh J. Changlani and Eli Chertkov and Eduardo H. da Silva Neto and Luca Delacretaz and Ismail El Baggari and G. M. Ferguson and William J. Gannon and Sayed Ali Akbar Ghorashi and Berit H. Goodge and Olga Goulko and G. Grissonnanche and Alannah Hallas and Ian M. Hayes and Yu He and Edwin W. Huang and Anshul Kogar and Divine Kumah and Jong Yeon Lee and A. Legros and Fahad Mahmood and Yulia Maximenko and Nick Pellatz and Hryhoriy Polshyn and Tarapada Sarkar and Allen Scheie and Kyle L. Seyler and Zhenzhong Shi and Brian Skinner and Lucia Steinke and K. Thirunavukkuarasu and Thaís Victa Trevisan and Michael Vogl and Pavel A. Volkov and Yao Wang and Yishu Wang and Di Wei and Kaya Wei and Shuolong Yang and Xian Zhang and Ya-Hui Zhang and Liuyan Zhao and Alfred Zong},
	journal={SciPost Phys. Comm. Rep.},
	pages={8},
	year={2025},
	publisher={SciPost},
	doi={10.21468/SciPostPhysCommRep.8},
	url={https://scipost.org/10.21468/SciPostPhysCommRep.8},
}

@article{osti_1262381,
title = {Data retrieved from the Materials Project for {Gd}$_2${Ru}$_2${O}$_7$ (mp-505216) from database version v2026.04.13.},
abstractNote = {Gd2Ru2O7 crystallizes in the cubic Fd-3m space group. The structure is three-dimensional. Gd3+ is bonded in a distorted body-centered cubic geometry to eight O2- atoms. There are two shorter (2.23 Å) and six longer (2.51 Å) Gd–O bond lengths. Ru4+ is bonded to six equivalent O2- atoms to form corner-sharing RuO6 octahedra. The corner-sharing octahedral tilt angles are 50°. All Ru–O bond lengths are 2.01 Å. There are two inequivalent O2- sites. In the first O2- site, O2- is bonded to two equivalent Gd3+ and two equivalent Ru4+ atoms to form a mixture of distorted corner and edge-sharing OGd2Ru2 tetrahedra. In the second O2- site, O2- is bonded to four equivalent Gd3+ atoms to form a mixture of corner and edge-sharing OGd4 tetrahedra.},
doi = {10.17188/1262381},
journal = {},
place = {United States},
year = {2020},
month = {7}
}

@article{CASTRO2021413227,
title = {{Effect of Mo substitution on the structure and electrical properties of Gd2Ru2O7 pyrochlore}},
journal = {Physica B: Condensed Matter},
volume = {619},
pages = {413227},
year = {2021},
issn = {0921-4526},
doi = {https://doi.org/10.1016/j.physb.2021.413227},
author = {A.A. Castro and J.L. Rosas-Huerta and R. Escamilla}
}

@article{Sibille2020, 
title={A quantum liquid of magnetic octupoles on the pyrochlore lattice},
volume={16},
DOI={10.1038/s41567-020-0827-7},
journal={Nature Physics},
author={Sibille, Romain and Gauthier, Nicolas and Lhotel, Elsa and Porée, Victor and Pomjakushin, Vladimir and Ewings, Russell A. and Perring, Toby G. and Ollivier, Jacques and Wildes, Andrew and Ritter, Clemens and Hansen, Thomas C. and Keen, David A. and Nilsen, Gøran J. and Keller, Lukas and Petit, Sylvain and Fennell, Tom},
year={2020},
pages={546–552}
}

@article{poree2025,
title = {{Dipolar-octupolar correlations and hierarchy of exchange interactions in ${\mathrm{Ce}}_{2}{\mathrm{Hf}}_{2}{\mathrm{O}}_{7}$}},
author = {Por\'ee, Victor and Bhardwaj, Anish and Lhotel, Elsa and Petit, Sylvain and Gauthier, Nicolas and Yan, Han and Pomjakushin, Vladimir and Ollivier, Jacques and Quilliam, Jeffrey A. and Nevidomskyy, Andriy H. and Changlani, Hitesh J. and Sibille, Romain},
journal = {Phys. Rev. B},
volume = {112},
issue = {18},
pages = {L180404},
numpages = {7},
year = {2025},
doi = {10.1103/j451-ztvr},
}

@article{Patri2020,
title = {Theory of magnetostriction for multipolar quantum spin ice in pyrochlore materials},
author = {Patri, Adarsh S. and Hosoi, Masashi and Lee, SungBin and Kim, Yong Baek},
journal = {Phys. Rev. Res.},
volume = {2},
issue = {3},
pages = {033015},
numpages = {18},
year = {2020},
doi = {10.1103/PhysRevResearch.2.033015},
}

@article{Hayami2024,
author = {Hayami, Satoru and Kusunose, Hiroaki},
title = {{Unified Description of Electronic Orderings and Cross Correlations by Complete Multipole Representation}},
journal = {Journal of the Physical Society of Japan},
volume = {93},
number = {7},
pages = {072001},
year = {2024},
doi = {10.7566/JPSJ.93.072001},
}

@article{Sutcliffe2026,
  author  = {Sutcliffe, Ruairidh and Hart, Kathleen and Chaudhary, Swati and Paramekanti, Arun},
  title   = {Pseudochiral Phonons from Octupolar Magnetic Order},
  journal = {Phys. Rev. Lett.},
  volume  = {137},
  pages   = {036703},
  year    = {2026},
  doi     = {10.1103/n11w-csdh}
}

@article{Watanabe2025,
  title = {{Dual-Circular Raman Optical Activity of Axial Multipolar Order}},
  author = {Watanabe, Hikaru and Oiwa, Rikuto and Mori, Hitoshi and Arita, Ryotaro},
  journal = {Phys. Rev. Lett.},
  volume = {137},
  issue = {5},
  pages = {056903},
  numpages = {9},
  year = {2026},
  month = {Jul},
  publisher = {American Physical Society},
  doi = {10.1103/m2dl-ghbf},
  url = {https://link.aps.org/doi/10.1103/m2dl-ghbf}
}

@article{Usachev2026,
  title = {{Nonlinear Optical Probing of Ferroic Octupolar Order Parameter in Collinear Altermagnet}},
  author = {Usachev, P. A. and Pisarev, R. V. and Pavlov, V. V.},
  journal = {Phys. Rev. Lett.},
  volume = {137},
  issue = {12},
  pages = {126902},
  numpages = {8},
  year = {2026},
  month = {Sep},
  publisher = {American Physical Society},
  doi = {10.1103/h1ht-vsrn},
  url = {https://link.aps.org/doi/10.1103/h1ht-vsrn}
}

@article{Banerjee2026,
  author        = {Banerjee, Saikat and Steinhofel, Tara and Lange, Florian and Eschrig, Matthias and Fehske, Holger},
  title         = {Light-driven octupolar inverse {F}araday effect and multipolar order in {M}ott insulators},
  journal       = {arXiv preprint},
  year          = {2026},
  eprint        = {2605.08049},
  archivePrefix = {arXiv},
  primaryClass  = {cond-mat.str-el}
}

@misc{Suganuma2026,
      title={{Electric toroidal octupolar symmetry in pyrite FeS$_2$ probed by Raman optical activity}}, 
      author={Yuki Suganuma and Gakuto Kusuno and Hikaru Watanabe and Rikuto Oiwa and Hitoshi Mori and Ryotaro Arita and Takuya Satoh},
      year={2026},
      eprint={2603.21756},
      archivePrefix={arXiv},
      primaryClass={cond-mat.mtrl-sci},
      url={https://arxiv.org/abs/2603.21756}, 
}

@article{Gao2026,
  title = {{Spectroscopic Demarcation of Emergent Photons and Spinons in a Dipolar-Octupolar Quantum Spin Liquid}},
  author = {Gao, Bin and Zhou, Zhengbang and Zhang, Tingjun and Podlesnyak, Andrey and Cheong, Sang-Wook and Kim, Yong Baek and Dai, Pengcheng},
  journal = {Phys. Rev. Lett.},
  volume = {136},
  issue = {25},
  pages = {256703},
  numpages = {7},
  year = {2026},
  month = {Jun},
  publisher = {American Physical Society},
  doi = {10.1103/svt2-m3pp},
  url = {https://link.aps.org/doi/10.1103/svt2-m3pp}
}

@article{Desrochers2024,
  author  = {Desrochers, F{\'e}lix and Kim, Yong Baek},
  title   = {Spectroscopic Signatures of Fractionalization in Octupolar Quantum Spin Ice},
  journal = {Phys. Rev. Lett.},
  volume  = {132},
  pages   = {066502},
  year    = {2024},
  doi     = {10.1103/PhysRevLett.132.066502}
}

@article{Amelin2020,
  author  = {Amelin, K. and Alexanian, Y. and Nagel, U. and R{\~o}{\~o}m, T.
             and Robert, J. and Debray, J. and Simonet, V. and Decorse, C.
             and Wang, Z. and Ballou, R. and Constable, E. and de Brion, S.},
  title   = {Terahertz magneto-optical investigation of quadrupolar spin-lattice effects in magnetically frustrated {Tb}$_2${Ti}$_2${O}$_7$},
  journal = {Phys. Rev. B},
  volume  = {102},
  pages   = {134428},
  year    = {2020},
  doi     = {10.1103/PhysRevB.102.134428}
}

\clearpage
\onecolumngrid


\renewcommand{\thesection}{S\arabic{section}}   
\renewcommand{\thetable}{S\arabic{table}}   
\renewcommand{\thefigure}{S\arabic{figure}}
\renewcommand{\theequation}{S\arabic{equation}}

\renewcommand{\figurename}{Fig.}
\renewcommand{\tablename}{Table}

\setcounter{equation}{0}
\setcounter{figure}{0}
\section*{Supplementary Material}

\section{Sample synthesis and characterization}

The measurements reported in the main text were performed on a pressed powder pellet 527~$\upmu$m-thick and 3~mm in diameter. The sample was synthesized by a conventional solid-state reaction. High-purity Gd$_2$O$_3$ and RuO$_2$ oxide powders, obtained from Sigma-Aldrich with 99.9\% purity, were mixed in stoichiometric amounts to yield 2 mmol of Gd$_2$Ru$_2$O$_7$. The mass of the precursor oxides were measured using a high precision analytical balance, after pre sintering for 600\textdegree C during 3 hours, reducing adsorbed moisture. The powders were then grinded on an agate mortar for about one and a half hours, the resulting mixture compressed into a pellet and placed under heat treatment for 1000\textdegree C during 24 h. The whole process was repeated and five thermal treatments were performed.

X-ray diffraction (XRD) data was acquired using a Bruker D8 advance diffractometer, from 10$^\circ$ to 70$^\circ$  using a 0.02$^\circ$  per step with a counting time of three seconds per point. Rietveld refinement was performed using the GSAS-II software, and the result is shown in Figure S1. The structure is compatible with the pyrochlore structure, and no precursor oxides nor impurities were detected within the sensitivity of the XRD measurement. The $\chi^2$ = 3.84, indicating good agreement with the pyrochlore structure. Rietveld refinement indicates an oxygen positional parameter x$_{48f}$ = 0.334, which lies within the expected range for pyrochlore oxides of 0.3125 $\leq$ x$_{48f}$ $\leq$ 0.375 and lattice parameter of a = 10.23 \AA, compatible with previous studies \cite{CASTRO2021413227}.

\begin{figure}[h]
    \centering
    \includegraphics[width=0.55\linewidth]{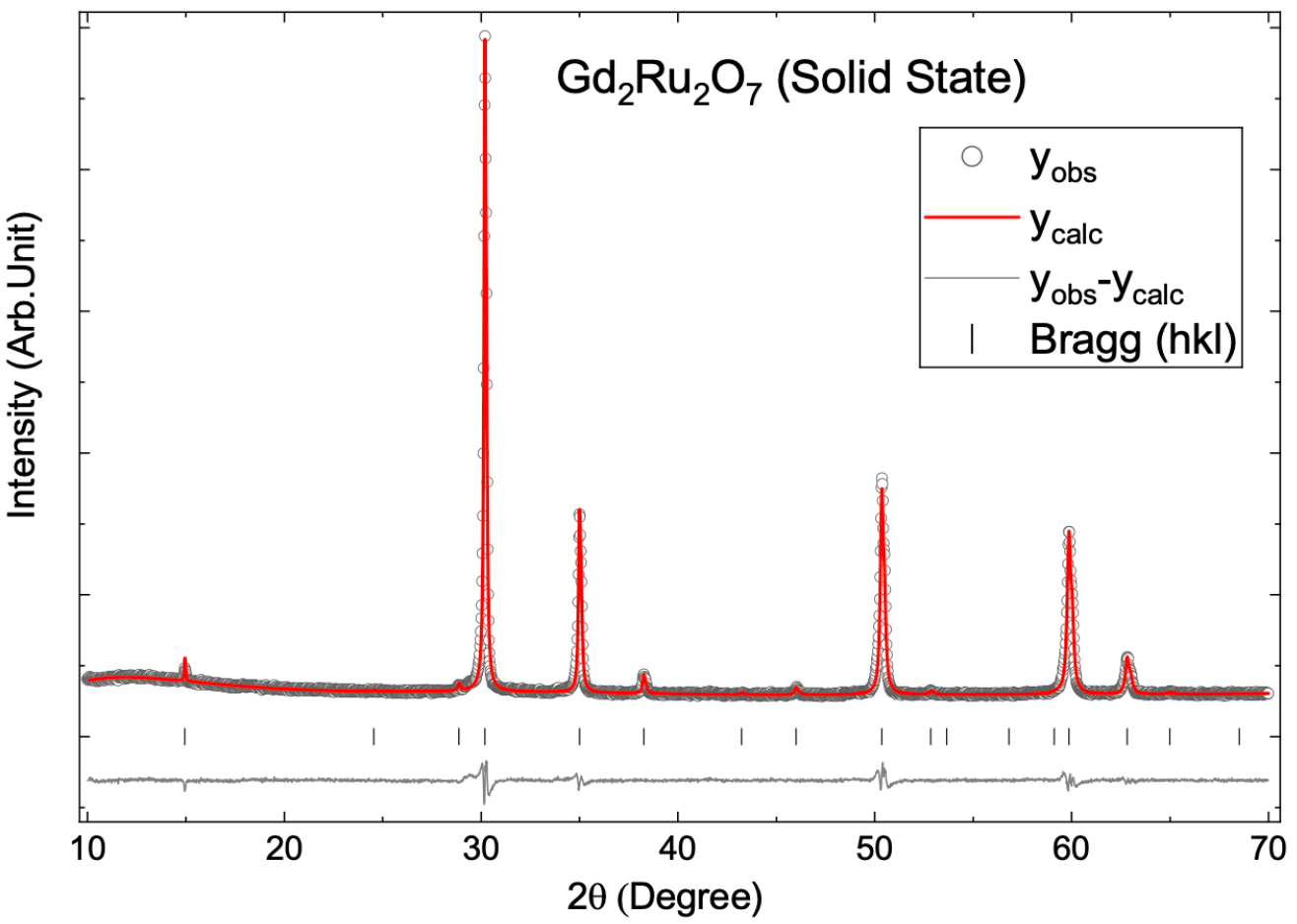}
    \caption{
    Gd$_2$Ru$_2$O$_7$ solid state sample XRD diffraction pattern, after five heat treatments of 1000\textdegree C for 24 h.}
    \label{fig4}
\end{figure}

\section{Experimental setup and data acquisition}

The terahertz time-domain spectroscopy (THz-TDS) experiment was conducted in transmission geometry using a commercial fiber-coupled photoconductive-antenna spectrometer (TeraFlash Pro, TOPTICA Photonics). The spectrometer was coupled to a cryofree superconducting magnet system (SpectromagPT, Oxford Instruments). Off-axis parabolic mirrors were used to focus the terahertz beam onto the sample, mounted over a 3-mm-diameter aperture, and to collect the transmitted radiation. Measurements were performed in Faraday geometry, with the terahertz propagation direction parallel to the external magnetic field, over the temperature range 1.6--200~K and for fields up to 7~T. The complex terahertz transmission was obtained as $T(\nu)=E_\textrm{sam}(\nu)/E_\textrm{ref}(\nu)$, where $E_\textrm{sam}(\nu)$ and $E_\textrm{ref}(\nu)$ are the Fourier-transformed sample and reference electric fields, respectively. A sample-free reference was acquired at $B_\textrm{ext}=$ 0~T for each temperature. Following the signal processing procedure described in Ref.~\cite{windowing}, the complex optical conductivity $\sigma(\nu)=\sigma_1(\nu)+i\sigma_2(\nu)$ was obtained through Fresnel inversion for a plane-parallel slab~\cite{Lloyd-Hughes2012-es}. The analysis presented in the main text focuses on the dissipative component $\sigma_1(\nu)$, which directly displays the absorptive resonances and was used to extract their oscillator strengths, resonance frequencies, and scattering times. Although both components were obtained from the complex transmission, the strongest resonances approach the transmission noise floor, making $\sigma_2(\nu)$ more sensitive to small phase uncertainties that can distort the corresponding dispersive line shapes.

\section{Lorentz-Oscillator Analysis}
To quantify the evolution of the magnetic and phonon resonances, the real optical conductivity spectra were fitted independently at each temperature and magnetic field. The fits were performed separately in the magnetic-excitation region below 0.5~THz and the optical-phonon region between 0.5 and 2.0~THz. In each spectral region, the resonance of interest was modeled as an uncoupled Lorentz oscillator superimposed on a smooth quadratic background,
\begin{equation}
    \sigma_1(\nu) = \mathrm{Re}\left[
        \frac{-i2\pi\nu\varepsilon_0\Omega_k^2}{\nu_k^2-\nu^2-i\nu/\tau_k}
    \right] + a_2\nu^2 + a_1\nu + a_0 .
\end{equation}
where $k=$ m and $k=$ p denote the magnetic excitation and optical phonon, respectively. The parameters $\Omega_k$, $\nu_k$, and $\tau_k$ correspond to the oscillator strength, resonance frequency, and scattering time, while $a_2$, $a_1$, and $a_0$ are the coefficients of the quadratic background. All parameters were allowed to vary independently for each spectrum. Figure~S2 presents representative fits for both spectral regions at selected temperatures and magnetic fields.

\vfill
\begin{figure}[h]
    \centering
    \includegraphics{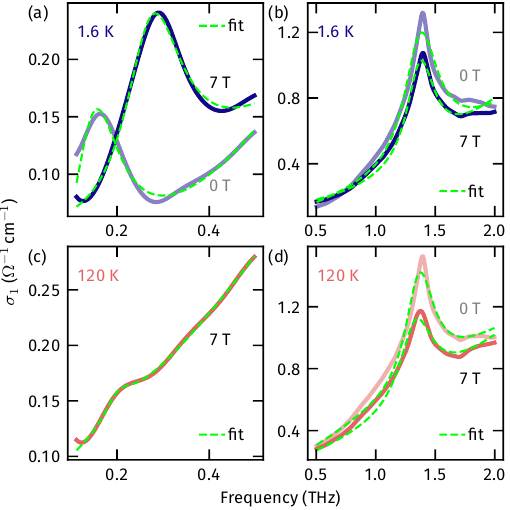}
    \caption{Representative fits to the real optical conductivity $\sigma_1(\nu)$ in the magnetic-excitation region below 0.5~THz [(a),(c)] and the optical-phonon region between 0.5 and 2.0~THz [(b),(d)] at 1.6~K [(a),(b)] and 120~K [(c),(d)]. Solid lines show the experimental spectra at the indicated external magnetic fields, and dashed lime lines show the fits using Eq.~(S1).}
    \label{fig_fit}
\end{figure}
\vfill
\vfill

\clearpage
\section{Integrated Spectral Weight}

To verify that the trends identified from the Lorentz-oscillator analysis do not depend on the assumed line shape, we also calculated the integrated spectral weight directly from the measured real optical conductivity,
\begin{equation}
W_k = \int_{\omega_{1,k}}^{\omega_{2,k}}\sigma_1(\omega)\,\mathrm{d}\omega = 2\pi\int_{\nu_{1,k}}^{\nu_{2,k}}\sigma_1(\nu)\,\mathrm{d}\nu ,
\end{equation}
where $k=$ m and $k=$ p denote the magnetic and phonon spectral regions, respectively. Fixed integration intervals of 0.1--0.5~THz for the magnetic region and 1.0--1.7~THz for the phonon region were used for all temperatures and external magnetic fields. No background subtraction was applied, so $W_k$ represents the total spectral weight within each selected interval rather than the isolated contribution of the corresponding Lorentz oscillator. As shown in Fig.~S3, the normalized integrated spectral weights reproduce the characteristic temperature scales and field-dependent trends identified from the fitted resonance parameters in the main text, confirming that these features are present directly in the measured spectra and do not arise from the fitting procedure.

\begin{figure}[h]
    \centering
    \includegraphics{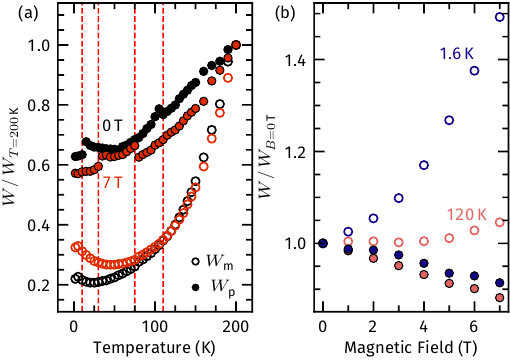}
    \caption{Normalized integrated spectral weights in the magnetic $W_\textrm{m}$, and phonon $W_\textrm{p}$ spectral regions as functions of (a) temperature at $B_\textrm{ext}=$ 0 and 7~T and (b) external magnetic field at 1.6 and 120~K. The integration intervals are 0.1--0.5~THz for $W_\textrm{m}$ and 1.0--1.7~THz for $W_\textrm{p}$. The temperature-dependent values are normalized by their respective values at 200~K, whereas the field-dependent values are normalized by their respective zero-field values. Vertical dashed lines in (a) indicate the characteristic temperatures discussed in the main text.}
    \label{fig4}
\end{figure}

\end{document}